\documentclass{aastex6}

\usepackage{epsfig, natbib} 
\usepackage[utf8]{inputenc}
\usepackage[T1]{fontenc}

\newcommand{\agn}{AGN}
\newcommand{\fesc}{$f_{esc}$}
\newcommand{\ha}{H$\alpha$}
\newcommand{\hi}{{\sc Hi}}
\newcommand{\hii}{{\sc Hii}}
\newcommand{\heii}{He {\sc ii}}

\newcommand{\ism}{ISM}
\newcommand{\lya}{Ly$\alpha$}
\newcommand{\lyc}{LyC}
\newcommand{\oii}{[{\sc Oii}]}
\newcommand{\oiii}{[{\sc Oiii}]}
\newcommand{\oiil}{[{\sc Oii}]$\lambda$3727}
\newcommand{\oiiil}{[{\sc Oiii}]$\lambda$5007}
\newcommand{\sii}{[{\sc Sii}]}

\slugcomment{manuscript revised 2017 August 22}

\begin{document}
\title{Haro 11: Where is the Lyman continuum source?}

\author{Ryan P. Keenan}
\affil{Department of Astronomy, University of Michigan, 1085 South University Ave., Ann Arbor, MI 48109}

\author{M. S. Oey}
\affil{Department of Astronomy, University of Michigan, 1085 South University Ave., Ann Arbor, MI 48109}

\author{Anne E. Jaskot}
\affil{Department of Astronomy, Smith College, Northampton, MA 01063}

\and

\author{Bethan L. James}
\affil{Space Telescope Science Institute, 3700 San Martin Dr., Baltimore, MD 21218}

\begin{abstract} 
Identifying the mechanism by which high energy Lyman continuum (LyC)
photons escaped from early galaxies is one of the 
most pressing questions in cosmic evolution.
Haro 11 is the best known local LyC leaking galaxy, providing
an important opportunity to test our understanding of LyC
escape.  The observed LyC emission in this galaxy
presumably originates from one of the
three bright, photoionizing knots known as A, B, and C.
It is known that Knot C has strong \lya\ emission, and
Knot B hosts an unusually bright ultraluminous X-ray source,
which may be a low-luminosity AGN.  To clarify the LyC source, we carry out
ionization-parameter mapping (IPM) by obtaining
narrow-band imaging from the {\sl Hubble Space
  Telescope} WFC3 and ACS cameras to construct spatially resolved
ratio maps of \oiii/\oii\ emission from the galaxy.  IPM traces the
ionization structure of the interstellar medium and allows us to
identify optically thin regions.  To optimize the continuum
subtraction, we introduce a new method for determining the best
continuum scale factor derived from the mode of the continuum-subtracted, image flux distribution.
We find no
conclusive evidence of \lyc\ escape from Knots B or C, but instead, we
identify a high-ionization region extending over at least 1 kpc from Knot A.
Knot A shows evidence of an extremely young age
($\lesssim 1$ Myr), perhaps containing very massive stars ($>100$ M$_\odot$).  
It is weak in \lya, so if it is confirmed as the \lyc\ source, our results
imply that \lyc\ emission may be independent of \lya\ emission.
\end{abstract}



\section{Introduction} \label{sec:intro}

The dominant source of the high energy Lyman continuum (\lyc) photons
responsible for reionizing the intergalactic medium (IGM) by $z\sim
5$ is one of the most important unsolved questions of cosmic evolution.
The principal candidates are photoionization by active galactic nuclei (\agn) 
or massive stars, but so far no conclusive evidence has been presented for either
model.  For example, \citet{fontanot12} find that the observed number of AGN is too
small to fully account for reionization, while \citet{fontanot14} find
that starburst galaxies can only be responsible if escape fractions (\fesc) of $\sim0.1-0.3$ are
assumed.   

Models suggest that mechanical feedback in starburst galaxies should
create optically thin paths through the \ism\ allowing high
fractions of the \lyc\ radiation to escape
\citep[e.g.,][]{clarke02, fujita03}.  Empirical studies challenge this picture,
finding relatively small \fesc\ for
\lyc\ radiation in both local and high-redshift starburst galaxies \citep{rutkowski16,siana15,leitet13}, 
but recently, LyC detections have been reported with \fesc\ as high as
0.13 from the low-mass, Green Pea galaxies at $z\sim 0.3$ \citep{izotov16a,izotov16b}.  
\lyc\ escape may not occur in all parts of a galaxy, nor is the radiation likely to be
emitted isotropically.  If ionizing radiation only leaves the galaxy through a
small solid angle, many instances of \lyc\ escape may not be oriented
along our line of sight, and so \lyc\ emission will not be
directly detectable.  This line-of-sight issue could
create a bias toward non-detection in direct searches for
\lyc-emitting galaxies, necessitating alternative methods of
detection \citep{zastrow13}. 

High-$z$ galaxies and the LyC-emitting Green Pea galaxies cannot be
resolved in enough detail for spatial mapping, so nearby LyC emitters
(LCEs) serve as our best opportunities to 
understand the process of \lyc\ escape.  Haro 11 is the best known
LCE in the local universe, having been the first to be identifed
\citep{bergvall06}, and it has been extensively studied
in many wavelengths.  Haro 11 is at redshift $z\sim0.021$
\citep{bergvall00}, which corresponds to a distance of 82.3 Mpc and a
scale of 0.384 kpc/arcsec, and it has a low metallicity of $Z\sim 0.2
Z_\odot$ \citep{james13}.  
Its kinematics strongly imply that it was formed by
the merging of two systems \citep{ostlin15}, which triggered a burst of
star formation starting $\sim$ 40 Myr ago \citep{adamo10,
  ostlin01}.  The star formation concentrates into three separate knots
identified as A, B, and C by \citet{vader93}, which are shown in Figure 1.  
Studies of Haro 11's star
clusters find that nearly all have formed in the past 40 Myr, with 60\% in
the last 10 Myr \citep{adamo10}; IFU spectra reveal 
Wolf-Rayet (WR) features in Knots A and B and suggest a recently ended WR
phase in Knot C \citep{james13, bergvall02}.  The low-metallicity,
chaotic environment, and high level of star formation are similar to
expectations for galaxies in the early universe thought to have driven
reionization.  
Haro 11 has a \lyc\ escape fraction of $3.3\pm
0.7$\% \citep{leitet11}.  However, the location of the \lyc\ source
within the galaxy remains unknown, and at least two of the knots have known
properties suggesting they could be the dominant \lyc\ source. 
While small compared to the escape fraction of some recently
discovered \lyc\ leakers \citep[e.g.,][]{izotov16a}, a $3.3\%$ escape fraction is 
significant and Haro 11's proximity offers an exciting opportunity for study.

\begin{figure} \label{fig:haromap}
\plotone{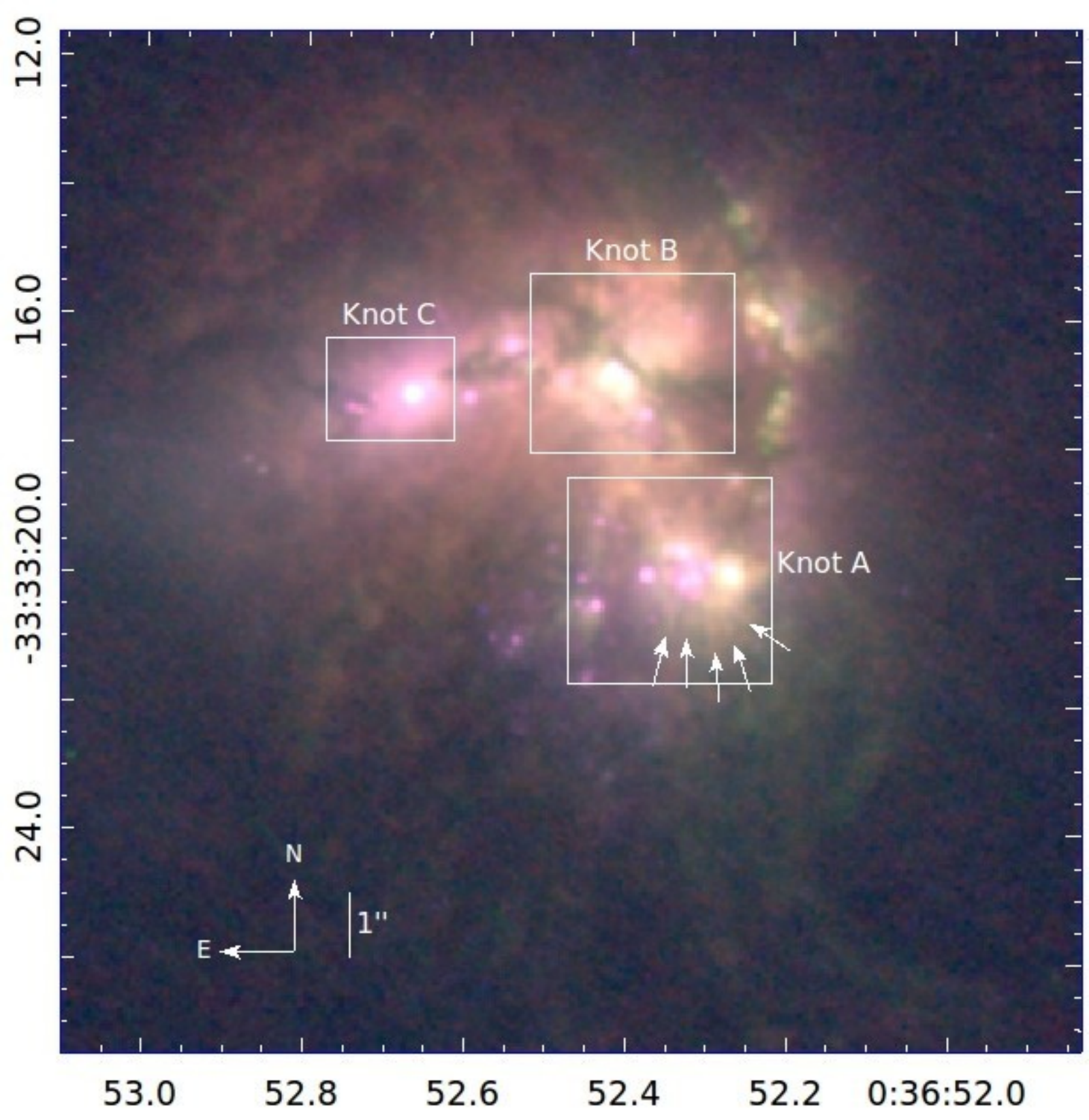}
\caption{Three-color image of Haro 11 in \oii\ (WFC3/FQ378N), \oiii\ (ACS/F505N),
  and $U$ (ACS/F336W), corresponding to red, green, and blue, respectively.
  The three major star forming knots are labeled.  White boxes correspond to subregions 
  for each Knot used in calculating the scaling factor.
  The arrows indicate finger-like structures projecting from
    Knot A, suggesting a Giuliani instability.  The scale bar shows
$1\arcsec$, corresponding to 0.384 kpc.}
\end{figure}

Knot C has generally been considered the strongest candidate LCE since it
is a strong \lya\ source \citep{kunth03,ostlin09,hayes15}.  It is 
usually assumed that LCEs are also strong \lya\ emitters since
both phenomena require low \hi\ column densities.  Knot B hosts an 
extremely bright ultraluminous X-ray source (ULX) dubbed Haro 11 X-1 by \citet{prestwich15}.  
They suggest that this source could be a low luminosity AGN (LLAGN),
which is especially interesting given 
results suggesting that faint AGN could produce sufficient energy to
explain cosmic reionization \citep[e.g.,][]{madau15,giallongo15}.  
However, there is no significant
\lya\ detection from Knot B \citep{ostlin09,hayes15}.  

Identifying which of the Knots is the LyC source in Haro 11 is therefore
essential to understanding the conditions responsible for LyC escape.  Can
we confirm the expected correlation between \lyc\ escape and
\lya\ escape, which would implicate Knot C?  Is the possible LLAGN
in Knot B responsible, implying that the dominant \lya\ source is
essentially unrelated to the \lyc\ source?  Or could the
\lyc\ emission originate from an entirely different location, in particular,
Knot A?  Since we cannot presently obtain spatially resolved imaging of
\lyc\ in Haro 11, we apply the technique of ionization-parameter mapping
\citep[IPM;][]{pellegrini12} to better understand the nebular ionization
structure of Haro 11 and locate its source of \lyc\ leakage.  


\section{Ionization-Parameter Mapping of Haro 11} \label{sec:method}

IPM entails constructing nebular emission-line ratio maps using two ions
with high and low ionization potentials, for example, \oiii/\sii.  
The boundary of optically thick regions can be identified by a layer
of emission in the low-ionization species, whereas optically thin
regions show emission only in the unbounded, high-ionization species.  
By locating the most optically thin regions,
IPM can provide indirect evidence for the presence of \lyc\ escape
features.  We caution that optically thick boundary regions with low
surface brightness can sometimes be difficult to detect, but IPM provides
quick insight for identifying areas that are strong candidates for LyC escape.
Our group has successfully used this method to identify candidate
optically thin H{\sc ii} regions in the Magellanic Clouds \citep{pellegrini12}, 
as well as such features in local starburst galaxies \citep{zastrow11,zastrow13}.  
Here, we apply IPM to Haro 11,
using the \oiiil\ line as our high-ionization
species and \oiil\ as our low-ionization species.  The choice of line pairs
from the same element controls for any element abundance
variations across the galaxy.  

We used the {\sl Hubble Space Telescope} ({\sl HST}) to obtain narrowband
imaging, using both the WFC3 and ACS cameras.  
To obtain the \oiil\ imaging, we used
WFC3 with the FQ378N filter, and for \oiiil, we used the ACS
FR505N ramp filter centered at 5110\AA\ to account for Haro 11's 
redshift of $z\sim0.021$ \citep{bergvall00}.  The \oii\ continuum image was obtained using
the WFC3 F336W filter.  For \oiii\ continuum an archival image in the
ACS F550M filter was used. In addition, we used archival imaging of
Haro 11 \ha\ emission in the FR656N ramp filter, centered at 6698\AA.  We
obtained a continuum image to pair with \ha\ using the WFC3 F763M
filter.  Total exposure times for WFC3 images were 8416 s in FQ378N, 1332 s in F336W, 
and 627 s in F763M; exposure times for ACS were 2379 s in FR505N, 471 s in F550M, 
and 680 s in FR656N. 

The STScI data pipeline provides flux-calibrated science images.  
Due to the sparseness of stars in our images, we used star clusters within the galaxy
to align the images.  These sources are unresolved and behave like point sources in our images.
We required that the centroids of point sources be within 0.3 pixels of one another in every image, 
and a closer match was usually obtained for each narrowband image and
its continuum counterpart.  

\subsection{A New Approach to Continuum Subtraction:  Mode of the flux
distribution} \label{sec:sub}

The narrowband images represent the sum of the line and
continuum emission, with the latter due primarily to diffuse stellar
background light.  Ordinarily, the continuum emission is removed by
subtracting an off-line image containing only continuum emission.  
This continuum image needs to be scaled by a scale factor $\mu$ to allow for
differences in filter band width, and more importantly, the variation
of the continuum spectral energy distribution (SED) between the two filters.  
Optimal scaling of the continuum filter flux is difficult to determine
and complicated by spatial variations in the stellar background
population, which have correspondingly varying SEDs.  
However, this issue is important
for IPM, since improper continuum subtraction in one or both
emission-line images can significantly affect the line ratios
calculated and the features that appear in ionization-parameter maps.  

Common methods for finding the scale factor $\mu$ include (1) using the
relative intensity of line-free objects or regions in the image, and (2) taking the ratio
of transmission efficiency for the filters.  \citet{hayes09} and
\citet{james16} use spatially resolved approaches in which (3) the
background SED is fitted for each image pixel, or binned pixels.  
The complications of these and other methods are reviewed by \citet{hayes09} and \citet{hong14}.  
Haro 11 is in a relatively sparse field and our images did not contain enough
sources outside the galaxy to apply approach (1).  Approach (2)
does not account for the continuum SED and
produced bad oversubtraction in parts of our images.  
Method (3) is heavily model-dependent and complicated to implement.  
We therefore seek a relatively simple, empirically based method to
apply to Haro 11.  \citet{hong14} describe a promising method for
determining the scaling factor using the skewness of the pixel
histogram of the continuum subtracted image as a function of 
$\mu$.  They find this relation to
be sensitive to the ratio of line to continuum emission in the line image.  
In particular, when the line-to-continuum ratio is high,
the optimal scaling is not well constrained by the skew.  Since we are
especially interested in regions at large galactocentric radius which are
dominated by such conditions, this motivates us to investigate other, similar
statistical methods for determining the best scaling. 
The use of a statistical measure on a large region of the image 
more directly addresses the goal of maximizing the number of pixels
for which a given $\mu$ is optimized.

The optimal continuum scale factor must be that which neutralizes
background continuum flux.  We experimented with using the
standard deviation $\sigma$, median, and mode of the
continuum-subtracted pixel values, in a spirit similar to Hong's et
al. use of the skew.
The standard deviation is that from the mean:
\begin{equation}
\sigma^2=\frac{\sum_{n=1}^N (x_n-X)^2}{N-1} \quad ,
\end{equation}
where $\sigma$ is the standard deviation, $N$ is the total number of pixels, $x_n$ is the flux
  in the $n$th pixel, and $X$ is the mean.
To calculate the median and mode, a histogram is created by binning the pixels into
bins of 0.5$\sigma$, which we found created large enough bins to accurately determine
the mode across the image without washing out variations introduced by
different values of $\mu$.  We used IRAF\footnote{IRAF is distributed
  by the National Optical Astronomy Observatory, which is operated by
  the Association of Universities for Research in Astronomy 
under cooperative agreement with the National Science Foundation.}
routines to find the median and mode.  The median is found by
integrating the histogram, and using interpolation to find the value
at which exactly half of the pixels are below and half above.
The mode is found by locating the maximum of the histogram, then fitting a peak by 
parabolic interpolation.  In the event that the first or last bin of the
histogram is the maximum value, then that bin is assigned as the mode.

\begin{figure}
\plotone{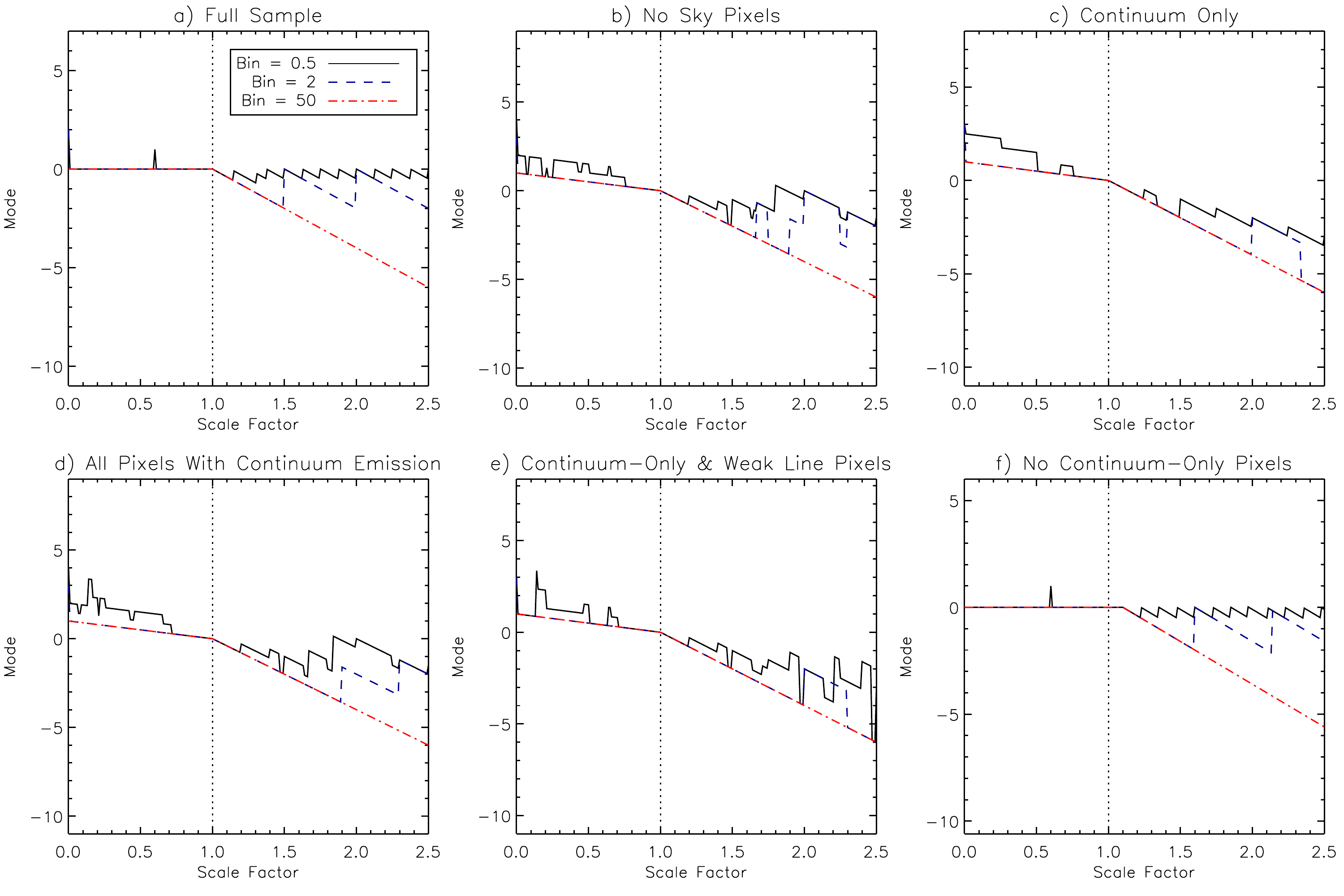}
\caption{The behavior of mode vs. $\mu$ for the fiducial synthetic
  dataset. The dashed vertical line shows the true scale factor of
  1. Bin sizes of 0.5, 2, and 50 are shown by solid black, dashed
  blue, and dot-dashed red lines, respectively. Each panel shows a
  different combination of pixels: (a) the full fiducial dataset,
  consisting of “sky” pixels with zero emission, continuum-only
  pixels, line-only pixels, and pixels with line-to-continuum ratios
  ranging from 0.1-10; (b) the fiducial dataset, excluding sky pixels;
  (c) only pixels with pure continuum emission; (d) continuum-only
  pixels and pixels with line-to-continuum ratios ranging from 0.1-10;
  (e) continuum-only pixels and pixels with a line-to-continuum ratio
  of 0.1; (f) the fiducial dataset, excluding the continuum-only
  pixels. \label{fig:jaskot}} 
\end{figure}

We find that the mode  of the image pixel
values has the most promising behavior.  The mode should 
decrease smoothly as the continuum is subtracted;
once the continuum becomes oversubtracted, negative pixels will cause
rebinning of the pixel histogram, inducing a change in the relation.
To explore the behavior of the mode with scale factor $\mu$, we generate an
artificial dataset consisting of background pixels with no emission,
and pixels with varying ratios of line and continuum emission. For our
fiducial model, we consider six categories containing four pixels
each: (i) background pixels with 0 counts each, (ii) continuum-only
pixels with continuum fluxes of 1-4 counts, (iii) diffuse
line-emitting pixels with line fluxes of 1-4 counts and no continuum
emission, (iv) pixels with continuum fluxes of 1-4 counts and a
line-to-continuum ratio of 0.1, (v) pixels with continuum fluxes of
1-4 counts and a line-to-continuum ratio of 1, and (vi) pixels with
continuum fluxes of 1-4 counts and a line-to-continuum ratio of 10. We
then vary the distribution of counts within each category, the
relative numbers of pixels in each category, the background level in
all pixels, and the bin size used to calculate the mode. We show
different combinations of these pixel categories and bin sizes in
Figure~\ref{fig:jaskot} and summarize the relevant results below. 

The optimal scale factor represents the point when the emission of the
continuum-only pixels exactly cancels out, reaching the background
level. These pixels start out with some initial distribution of flux
values. As $\mu$ is increased, their average flux and
mode value decrease. At the optimal scale factor, their flux values
pile up at the background level and then spread out again toward a
range of negative values as the pixels become increasingly
over-subtracted. Notably, the pixels that start out with the strongest
continuum fluxes show the most rapid decrease in flux with scale
factor. These pixels have the highest flux values before the optimal
scale factor and the most negative values after the optimal scale
factor.  Since flux
bins are defined starting with the lowest data values, the pixels with
the weakest continuum flux set the mode bin value just below the
transition point $\mu$ value, while the pixels with the strongest
continuum flux set the mode bin value above the transition. This
transition therefore corresponds to a change in slope in a plot of
mode vs. $\mu$, as seen in Figure~\ref{fig:jaskot}. 

The mode-based continuum subtraction method works best when the image
contains a large number of pixels with pure continuum emission (Figure
\ref{fig:jaskot}a--e). However, in cases where background pixels dominate the mode
value, this method is sensitive to even a single over-subtracted
pixel. Once a pixel’s flux decreases below the background level, it
defines the value of the lowest flux bin; since all other flux bins
are defined as integer numbers of bins after the first bin, the lowest
bin sets the lower and upper flux limits of all subsequent bins. The
first over-subtracted pixel can therefore redefine the value of the
bin that contains the background pixels and change the mode
value. Pixels containing both line and continuum emission behave
similarly to the pure continuum pixels but reach their
over-subtraction point at higher scale factor values (Figure
\ref{fig:jaskot}f). If no pure continuum pixels are present, the pixels with line
emission will generate the observed slope transition and the true
scale factor will be overestimated; the observed transition point
therefore sets an upper limit on the optimal scale factor. The mode
values for diffuse-line and background pixels do not depend on the
scale factor, because they have no continuum emission. When present,
these continuum-free pixels can set the initial mode value (e.g.,
Figure~\ref{fig:jaskot}a,f), but the exact bin that contains this mode value will
vary depending on the value of the lowest flux bin.  We again emphasize that
the mode-based method is sensitive to the first over-subtracted
pixels, including ones with spuriously low values.  However, the
optimal value should still induce a subsequent transition as well.
This method therefore requires empirical verification of the best
value for $\mu$.

Because the slope transition represents a change in the value of the
lowest flux bin, it occurs at the same scale factor regardless of the
bin size used to calculate the mode; however, the change is most
obvious with large bin sizes (Figure~\ref{fig:jaskot}). Small bin sizes may show
more discontinuities, as pixels abruptly shift into or out of a
bin. Nevertheless, the overall slope on either side of the
discontinuity remains constant and only changes at the optimal scale
factor. Oscillations typically appear after the optimal scale factor,
and their amplitude and frequency depend on bin size. As
over-subtracted pixels decrease in flux, they continually re-define
the value of the lowest flux bin and all higher flux bins. The value
of the flux bin that contains the mode will oscillate as the mode
changes from the highest value in its flux bin to the lowest value in
its flux bin. Larger bin sizes wash out discontinuities and
oscillations, so that the only remaining transition is the slope
change as over-subtraction begins.

\begin{figure} 
\plotone{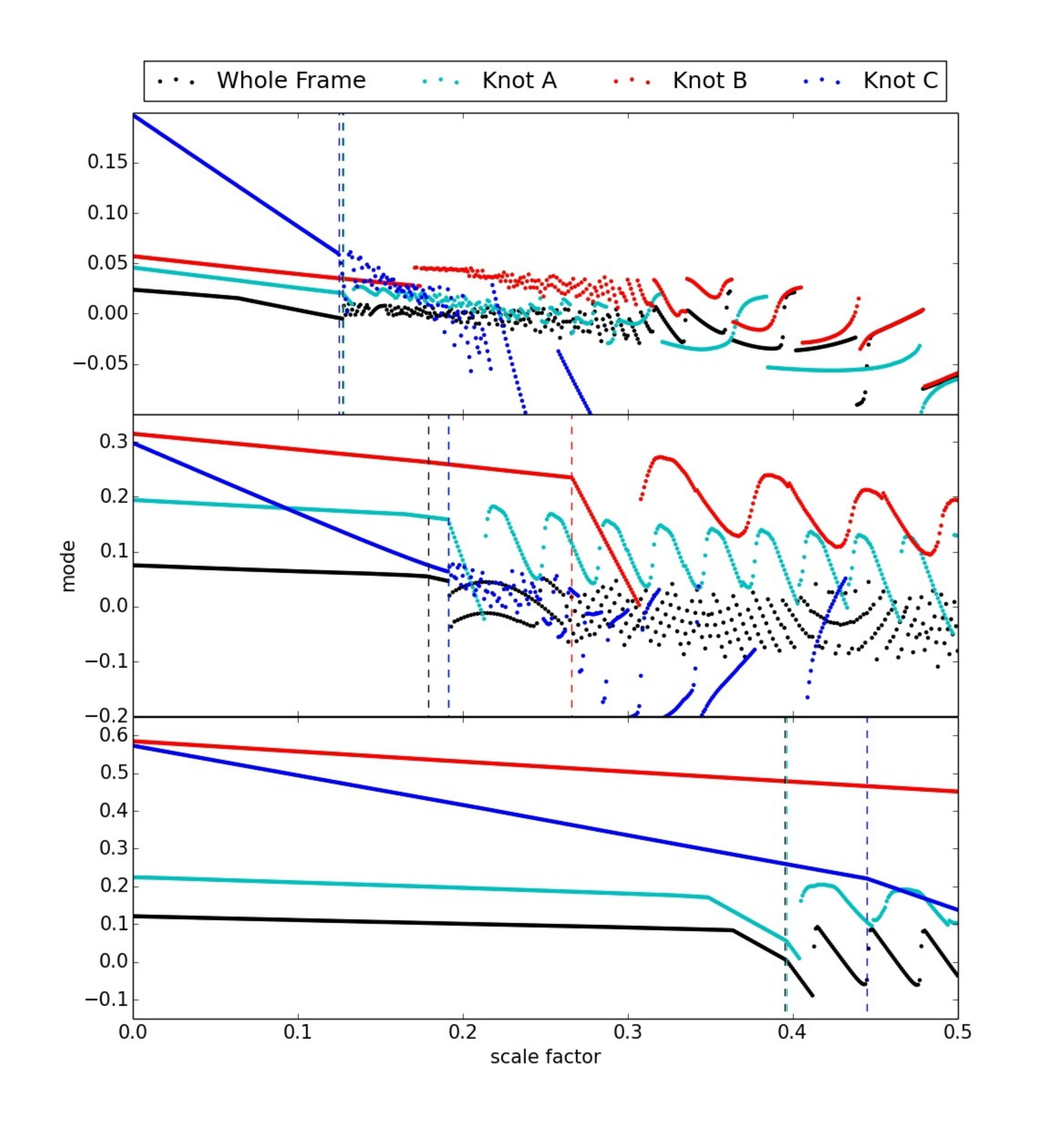}
\caption{Mode of continuum-subtracted pixel values as a function of
  scale factor $\mu$ for \oii\ (top), \oiii\ (middle) and
  \ha\ (bottom).  The whole frame and subregions defined in
  Figure~\ref{fig:haromap} are distinguished with the shown colors.
  The vertical dashed lines show the adopted $\mu$ values, using the
  same color scheme.
\label{fig:calib} }
\end{figure}

\begin{figure} 
\plotone{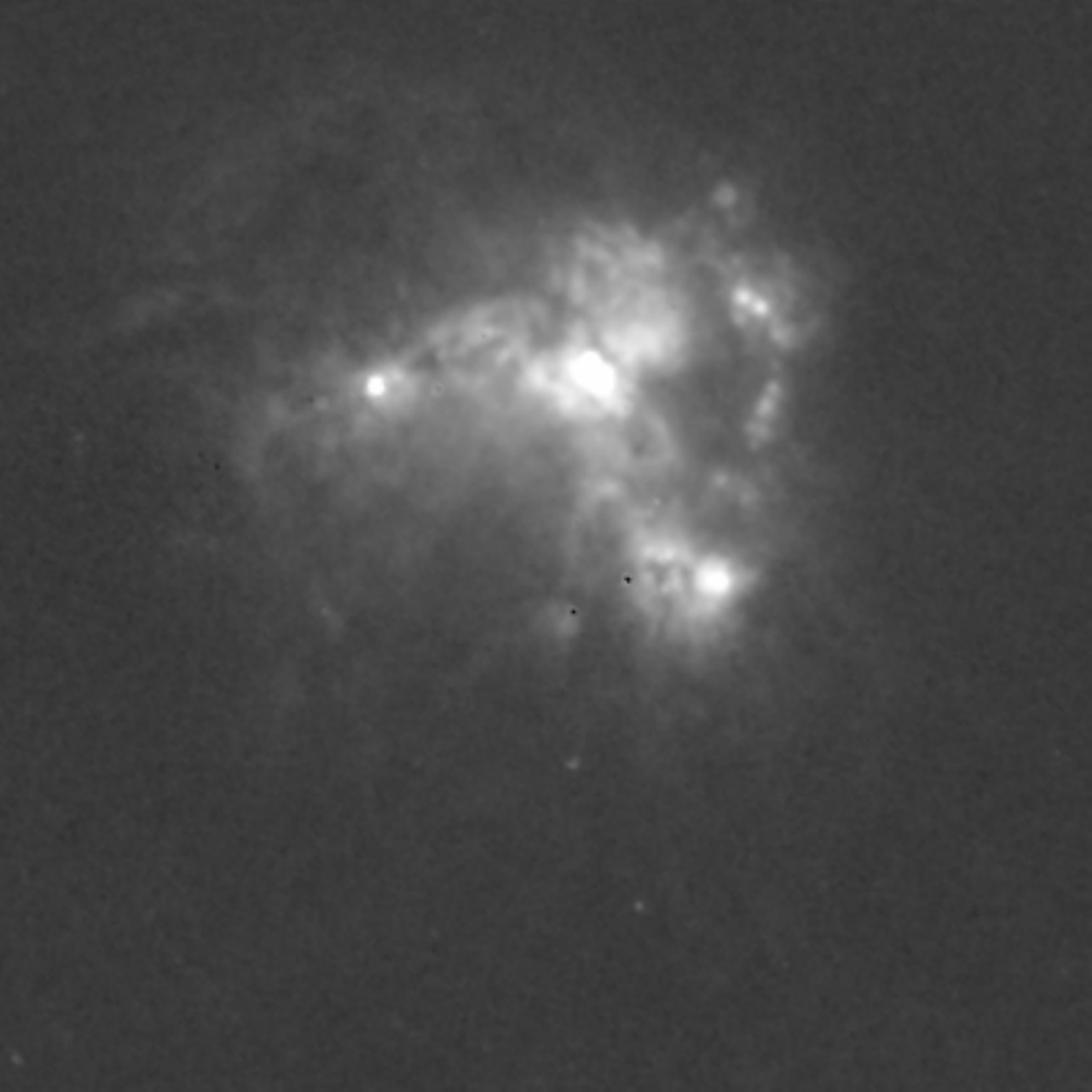}
\caption{Continuum-subtracted, emission-line image in \ha.
\label{fig:hal}}
\end{figure}

Figure~\ref{fig:calib} shows the mode as a function of scale factor for our imaging
in each line.  For the whole-frame images (black), we find clear breaks in
this relation for each line, which give scaling values of  
$\mu_{\rm WF}=0.127$ for \oii, $\mu_{\rm WF}=0.179$ for \oiii, and {$\mu_{\rm WF}=0.395$} for \ha.  
Figure~\ref{fig:hal} shows, as an example, the \ha\ image, continuum-subtracted using the mode method.
Here, we see an example where the second slope break, rather than the
one at the lowest $\mu$, corresponds to the optimal scale factor.
We also determine the scale factors for smaller regions around each
Knot, shown in Figure~\ref{fig:calib}, with Knot A in cyan, Knot B in red,  
and Knot C in blue.  The regions are shown by the white boxes in
Figure~\ref{fig:haromap}.  For Knot A, we find a good fit in all
bands, which generally agrees well with the
scale factor for the entire image.  We find $\mu_{A}=0.127$
for \oii, $\mu_{A}=0.191$ for \oiii, and $\mu_{A}=0.396$ for \ha.   
Determining the scale factor is most difficult in Knot B.  For
\oii, there is no obvious point at which the slope changes, although
around $\mu_B\sim 0.2$ would be an upper limit.
For \oiii\ we find a clear optimal value at
$\mu_{B}=0.266$.  For \ha, we find no break
below $\mu_{B}=0.500$.  The break in the mode for Knot B occurs at a
much higher value  than for the other regions, in all
bands.  This implies that the continuum is fainter in this region,
which may be related to the pronounced dust lane seen in Figure~\ref{fig:haromap}.
For Knot C, we find $\mu_{C}=0.125$ for \oii\ and $\mu_{C}=0.445$ for \ha.  
For \oiii, there is a clear break where noise begins at
$\mu_{C}=0.191$, consistent with values for other regions.
Comparison of the mode for the entire image, Knot A, Knot B, and Knot C 
shows that Knots A and C typically have very similar scaling, and agree 
with the scaling for the whole image.  

\begin{figure} 
\plotone{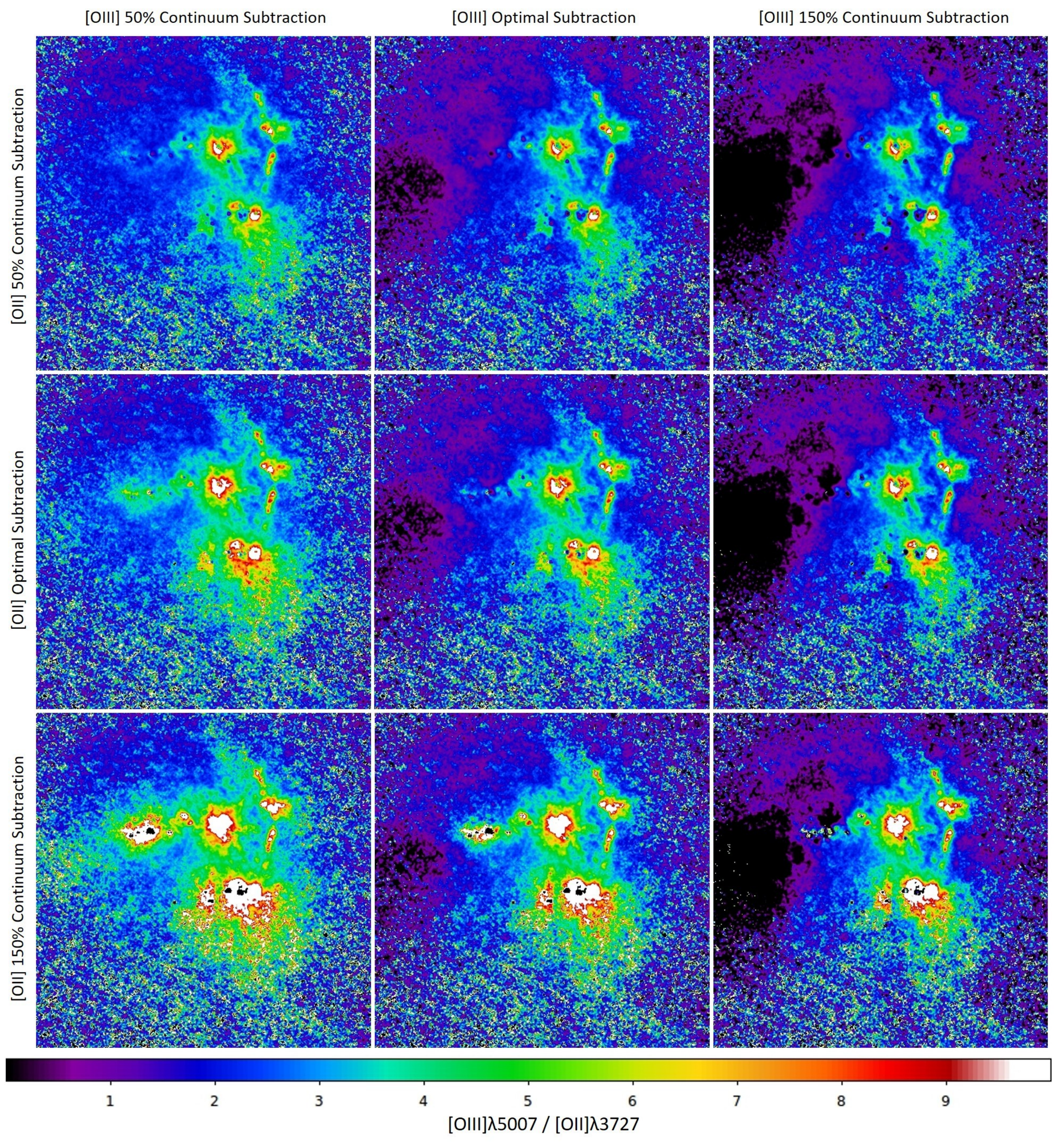}
\caption{Ratio maps of \oiii/\oii\ for different combinations
    of errors in continuum subtraction.  Maps in the left, center, and
    right columns are constructed with \oiii\ continuum subtraction of
    $0.5\times\mu_{\rm WF}$, $1.0\times\mu_{\rm WF}$, and $1.5\time\mu_{\rm WF}$,
    respectively. The upper and lower rows have \oii\ continuum
    subtraction at $0.5\times\mu_{\rm WF}$, $1.0\times\mu_{\rm WF}$, and $1.5\times\mu_{\rm WF}$,
    respectively.  The center panel shows the ratio map created using
    the optimal continuum subtraction in both bands.}
\label{fig:ratiomaps}
\end{figure}

Due to the close agreement between three of our four measurements, we chose to
use the scale factor determined for the whole frame to calibrate our
final \oiii/\oii\ ratio map, which is shown in the center
  panel of Figure~\ref{fig:ratiomaps}.  These values are
$\mu_{\rm WF}=0.127\pm0.002$ for \oii, $\mu_{\rm WF}=0.179\pm0.012$ 
for \oiii, and $\mu_{\rm WF}=0.395\pm0.050$
for \ha, where the errors are determined by the maximal difference between the 
scale factor for the full frame, Knots A and C.  These values produce images with 
few or no continuum pixels subtracted below the level of the background, and with greatly
decreased continuum flux compared to unsubtracted images.
Using a single scale factor for the whole galaxy ignores
spatial variations in the continuum SED, and Figure~\ref{fig:calib} shows that some
variation is present, as expected.  However, since we are mainly interested in
the \oiii/\oii\ ratio map at large galactocentric distances, this ratio
is fairly insensitive to the exact value of $\mu$, since the continuum has very low
emission, if any, in these regions.

\section{LyC Source Candidates} \label{sec:discussion}
\subsection{Ionization Parameter Mapping} \label{sec:discussion1}

The center panel of Figure~\ref{fig:ratiomaps} shows the optimally continuum-subtracted ratio map of 
\oiii/\oii\ emission.  There is a highly ionized region extending at
least 3$\arcsec$ ($> 1$ kpc) from
the center of Knot A into the outskirts of the galaxy to the south and southwest.  
The lack of a clear transition to low \oiii/\oii\ at the edge of this region
strongly suggests that it is optically thin and may be allowing the escape of
\lyc\ into the circumgalactic medium.  By contrast, the confined
morphology of \oiii/\oii\ for the rest of the galaxy, in particular,
around Knots B and C, suggests that these other regions are surrounded
by optically thick envelopes.  The detection of this highly ionized
region extending from Knot A is robust to errors in continuum
subtraction.  Figure~\ref{fig:ratiomaps} shows the ratio maps constructed from
\oiii\ images with the continuum subtracted at $0.5\times\mu_{WF}$ (left column), 
$1.0\times\mu_{WF}$ (center column), and $1.5\times\mu_{WF}$ (right column),
and \oii\ images subtracted at the same increments in the top, center,
and bottom rows, respectively.  Thus,
the ratio map with optimal subtraction in both images appears in the center panel.
We note that our analysis of the continuum scaling above (\S~\ref{sec:sub}) suggests an 
uncertainty in $\mu$ of less than a factor of 0.1 for both \oii\ and \oiii.
The extended, high-ionization region is present in all cases, despite notable changes in the
appearance of other regions.  It is especially noteworthy that Knot C,
which is usually assumed to
be the LyC-emitting region, generally appears to be optically thick in
the LyC, although it is optically thin in \lya\ \citep{kunth03}.  
Knot C does show an ionized region stretching to the east in
the case where we undersubtract the continuum in \oiii\ and
oversubtract in \oii\ (Figure~\ref{fig:ratiomaps}, lower left quadrant).  However, this feature only appears
when very large errors in continuum subtraction are assumed.

IPM in \oiii/\oii\ primarily maps the galaxy's ionization structure
in the plane of the sky.  However, assuming the \lyc\ detection from
Haro 11 is real, we should also expect at least one
ionized region to be optically thin to \lyc\ along the line of sight,
to explain previous direct detections of \lyc\ emission from Haro 11.  
The ratios of \oii\ and \oiii\ to \ha\ can be used as an indicator for
ionization in the line of sight \citep{pellegrini12}.  In particular,
a region is likely to be optically thin when it shows both a high
\oiii/\ha\ ratio and low \oii/\ha\ $\lesssim 0.1$.  
Weak \oii/\ha\ again indicates the lack of a nebular zone in the
low ionization species, whereas a comparatively high \oiii/\ha\ ratio in the
same line of sight then indicates that the gas is highly ionized.  
Figure 6 shows maps of these ratios.  Knot A indeed has both low \oii/\ha\ and high
\oiii/\ha, indicating it is likely optically thin in the line of
sight.  This further indicates that Knot A is a strong candidate for the galaxy's \lyc\ emission source.  
Knot B also shows extremely low values of \oii/\ha, but the excitation
is much lower in Knot B, as indicated by lower \oiii/\ha.  
Furthermore, as we have seen above, the continuum subtraction for \oii\ in Knot B is uncertain.
Dust could play a role in suppressing the \oii\ line emission relative to the 
redder \oiii\ line.  There is a noticeable dust lane through the region, which adds further
uncertainty to any conclusions about Knot B.  We also see no indication that
Knot C is optically thin in the line of sight.

\begin{figure}
\figurenum{6}
\plotone{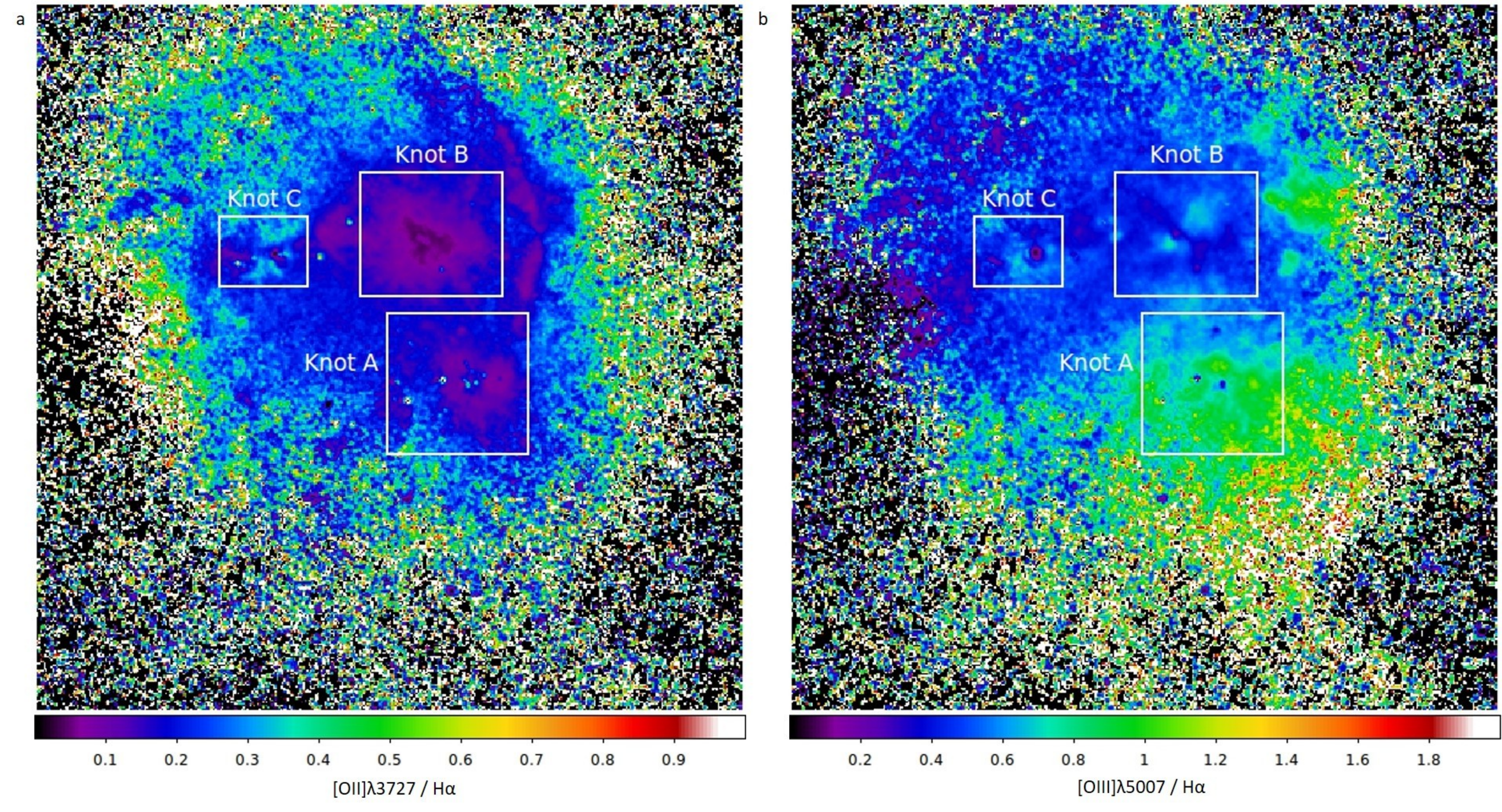}
\caption{Panel a: \oii/\ha\ ratio map.  Panel b: \oiii/\ha\ 
  ratio map.  The combination of strong \oiii\ and low \oii\ suggests
  Knot A as a source for direct detections of \lyc\ from Haro 11.  A
  small localization on Knot B is also a candidate.} 
\end{figure}

\subsection{A, B, or C? Implications for LyC Escape} \label{sec:discussion}

Of the three dominant sources in Haro 11, both Knots B and C have previous 
evidence suggesting that they could be the \lyc\ source(s).  
Knot C is visually the brightest knot and shows strong \lya\ emission
\citep{kunth03,ostlin09,hayes15}.  It appears to be the nucleus of one of the two galaxies in
this merging system \citep{james13,ostlin15}, and \citet{prestwich15} also identify 
Knot C as the host of a ULX source with an
X-ray luminosity of $L_X\sim5\times10^{40}$ erg s$^{-1}$, dubbed Haro 11
X-2.  They suggest that this source may provide enough mechanical power
to clear channels through the \ism\ promoting a low \lya\ optical depth.  
Such geometry would also enhance the escape of LyC radiation.  
However no previous work has been able to directly attribute \lyc\ emission to Knot C, 
and \citet{rivera-thorsen17} find that the column density of neutral gas in the Knot
is too high to allow \lyc\ escape from a density bounded ISM, although
they suggest it may be possible if the covering fraction of denser gas
is less than unity.
While we find that Knot C exhibits an ionization feature to the east
in one calibration of the continuum subtraction
(Figure~\ref{fig:ratiomaps}), our optimal subtraction shows Knot C to be in
a low ionization state compared to much of the galaxy, which does not
support it being the \lyc\ source.

In Knot B, \citet{prestwich15} report an unusually luminous ULX with 
$L_X\sim10^{41}$ ergs s$^{-1}$, which could be a low-luminosity AGN.  
Recent models have suggested that LLAGN could be important, or even vital,
for cosmic reionization \citep{fontanot12,madau15,giallongo15}.  
It should also be noted that LLAGN are now known to be relatively
common in dwarf galaxies \citep[e.g.,][]{reines13,lemons15,baldassare15}.
However, \citet{prestwich15} note that the source in Knot B is more likely to be a luminous
X-ray binary, and possibly a supermassive black hole progenitor. They show that
an X-ray binary could produce mechanical power comparable to the supernova
feedback for an entire cluster, indicating that such a system could clear paths
for \lyc\ escape.  Thus, Haro 11 may provide an important opportunity to evaluate AGN, 
or possibly black hole binaries, versus massive stars as cosmic \lyc\ sources.  
Figure 6 shows that Knot B itself exhibits a 
low ratio of \oii/\ha\ suggesting that it may be a \lyc\ leakage
source along the line of sight.  However, it is clearly
ionization-bounded transverse to our sightline in our
\oiii/\oii\ ratio maps, showing no signs of \lyc\ escape along the
plane of the sky.  Thus if Knot B is optically thin, it would be
consistent with suggestions  that \lyc\ leakage is not isotropic, but
would require an especially narrow ionization cone along 
the line of sight to show no hints in other directions \citep[e.g.,][]{zastrow11}.  
Our data cannot definitively rule out Knot B as the dominant
\lyc\ emitter in Haro 11.  

In contrast to the negative and ambiguous results for Knots C and B, respectively,
our ionization-parameter mapping indicates that Knot A may be
the strongest candidate for the \lyc\ emission.  The large, highly
ionized region extending to the outskirts of the galaxy implies that
the region southwest of Knot A is optically thin to \lyc\ and could
allow large escape fractions in those directions.  Preliminary results
presented by \citet{bik15} for \sii/\oiii\ ratios also show this feature.  
In addition, the low ratio of \oii/\ha\ for Knot A itself suggests
that this potentially optically thin region also extends along our line of sight.  
We have no reason to expect that large dust clouds could
systemically depress the \oii\ relative to \oiii\ flux in the galaxy outskirts.  
Additionally the observations by \citet{bik15} use tracers where the high
ionization species is the bluer line, ruling out reddening as an explanation.
Thus, we consider Knot A to be at least as good a candidate for the
origin of \lyc\ emission as Knots B and C, and likely even stronger.  

Previous studies have found \lyc-leaking candidates with narrow,
cone-shaped \lyc\ escape regions \citep{zastrow13, zastrow11}.  The
ionization in Knot A is over a broader region than the cones
reported in other galaxies, but still covers only a limited
directional range, adding further support to suggestions that
\lyc\ escape is highly nonuniform across regions of a galaxy.  The
growing number of galaxies with relatively narrow ionization regions
suggests that line of sight bias is a problem for direct
searches for galaxies with high \lyc\ escape fractions.  

Knot A also exhibits other noteworthy features.  In particular, Figures 1 and 3 show
remarkable, finger-like structures projecting southwards from the
knot.  These structures are startlingly similar to those seen in
simulations by \citet{freyer03} and \citet{garciasegura96}.  The
dominant effect is the \citet{giuliani79} instability, which is a
specific case of the \citet{vishniac83} thin shell instability applied
to ionization fronts.  The simulations show that ionized fingers form
at extremely early stages in \hii\ region evolution, on the order of
$10^5$ yr.  They occur when the instability causes the thin shell
driven by the ionization front to fragment, leading to recombination
in the remnant clumps, which cast shadows between the low-density,
ionized fingers.  The effect can be amplified by mechanical feedback
from a wind-driven shell \citep{freyer03}.  On a similar timescale,
the clumps then merge into a more amorphous configuration and are
swept up by the stellar wind shell, causing the fingers to disappear
again.  Thus, these ionized features are short-lived phenomenon that 
appear at a very early stage of the massive star feedback.

The features in Knot A appear to be examples of the
Giuliani instability.  The morphology of the fingers is uniform
and consistent with light rays emitted through shadowed regions, as
opposed to showing irregular surfaces of dense gas, as seen in
pillars.  This is consistent with the
radiation-dominated conditions indicated by the extreme ionization
seen in Figures 5 -- 6, and probable optically thin state, in the same
directions.  Finally, these directions lead directly out from the
galaxy into a lower density environment, which is again a condition
promoting the Giuliani effect.  Interestingly, the extremely young age,
$\lesssim 1$ Myr, is also implied in the nearby \lyc\ candidate Mrk~71
\citep{micheva17}.  The extreme Green Pea
galaxies, which are the best class of local LCE candidates,
also have young ages, in some cases $\lesssim 2$ Myr \citep{jaskot13}.

The suggestion of such an extremely young age for Knot A is at 
odds with the slightly more evolved age of 4.9 Myr reported by \citet{james13}.  
This value is driven by the presence of WR stars within the stellar 
population synthesis models; since WR stars form around 3 -- 5 Myr, 
the models yield corresponding instantaneous burst ages when 
WR features are detected.  However, recent work is pointing to 
the contribution of very massive stars (VMS) with masses 
$>100\ \rm M_\odot$ to WN features in the extremely young super 
star clusters 30 Doradus in the LMC \citep{crowther10,crowther16} 
and NGC 5253-5 \citep{smith16}.  The latter work shows that strong 
VMS candidates in NGC 5253-5 generate only broad, \heii\ 
$\lambda4686$ emission, with no stellar {\sc [N iii]} $\lambda\lambda 4634-4641$ 
contribution to the WR blue bump.  We note that the same is observed 
in Knots A and B of Haro 11, as seen in the FLAMES IFU spectra 
of \citet[][their Figure~10]{james13}.  NGC 5253-5 has an 
estimated age of $1\pm1$ Myr \citep{calzetti15}, and 30 Dor 
similarly has an age of $1.5^{+0.3}_{-0.7}$ Myr \citep{crowther16}.  
Both systems are believed to be optically thin in the LyC \citep{zastrow13,pellegrini12}.  
Thus Knot A may be similar to these newborn, optically thin, 
super star cluster systems.

If Knot A, rather than Knot C, turns out to be the dominant \lyc\ source in Haro
11, then this has major implications in understanding the
relationship between \lya\ and \lyc\ emission.  In general, it is
assumed that \lyc\ emitters are also \lya\ emitters, since \lya\ is
produced by the recombination of ionizing radiation, and the
optical depth of \lya\ is extremely sensitive to small \hi\ columns.
Thus, the conditions for \lya\ escape are also similar to those of
\lyc\ escape.  However, we note that the escape processes differ
substantially:  \lya\ is optically thick at an \hi\ column
$\log N$(\hi)$\sim13$ that is $10^4$ times lower than the threshold
for \lyc, $\log N$(\hi)$\sim17$.  Thus, in many situations, regions 
that are optically thin in \lyc\ will be optically thick to \lya.
However, \lya\ scatters strongly and may re-emitted
non-isotropically and far from the original line of sight.
  Much longer path lengths also enhance dust absorption of
  \lya\ relative to \lyc.
Thus, the relationship between these two escape processes
may be more complicated than is generally appreciated.  For Haro 11,
we find that, while this galaxy emits in both \lyc\ and \lya, the
respective sources may not coincide.  Spatially resolving the
\lyc\ emission in this galaxy will greatly clarify our understanding
of the radiative transfer.

\section{Conclusions} \label{sec:conclusion}

In summary, we use {\sl HST} WFC3 and ACS narrow-band imaging to carry
out ionization-parameter mapping in an attempt to clarify the spatial origin of the
\lyc\ emission in Haro~11.
We present a new statistical method for optimizing the continuum
  subtraction, based on the mode of the image flux distribution.
Applying this technique, we obtain continuum-subtracted,
emission-line images in \oiii, \oii, and \ha,
providing spatially resolved insight on  
the LyC escape in this known LyC-emitting galaxy.  
Haro~11 has strong \lya\ emission from Knot C and 
may host a LLAGN in Knot B,  which are both potential indicators of a LyC source.  
However, we find no evidence for LyC escape from Knot C,
consistent with recent results of \citet{rivera-thorsen17}.  Knot B
is likely optically thin to LyC along the line of sight, as indicated by a low \oii/\ha\ ratio
on the center of the Knot.  However, along the plane of the sky, Knot B shows a
smooth drop in \oiii/\oii, indicating that it is radiation-bounded.  

Instead, we find the strongest evidence for LyC escape to be from Knot A, which has
not previously been suggested as a candidate for the LyC source.  The
\oiii/\oii\ ratio maps show a highly ionized region extending at least 1~kpc 
from the center of the Knot into the surrounding medium, suggesting that it is optically 
thin to LyC radiation.  Additionally a low \oii/\ha\ ratio at the
center of Knot A implies that it is also optically thin along the line
of sight, further enhancing the possibility that it is the
source of directly detected LyC radiation.  We suggest that the observed 
broad \heii\ $\lambda 4686$ emission in Knot A may be due to 
VMS stars rather than classical WR stars, pointing to an extremely 
young age on the order of 1 Myr.  This is consistent with the appearance 
of ionized fingers that may be generated by the Giuliani instability.  
It is possible that the candidate knots each provide partial
contributions to the LyC emission, which suggests the possibility  
that a combination of LLAGN/accretion and stellar feedback may
generate \lyc\ leakage.  If Knot A is confirmed as the LyC
source, this result would have strong implications for our
understanding of the relationship between \lya\ and \lyc\ emission,
implying that although Knot C is a strong \lya\ emitter, it may be an insignificant \lyc\ emitter, 
while the opposite may be the case in Knot A. This would imply that 
\lya\ and \lyc\ emission might be substantially independent.
It would also provide further evidence that LyC escape is highly
anisotropic and therefore challenging to detect directly \citep[e.g.,][]{zastrow11}.  Further
work is needed to identify which of the Knots is the true source of
the observed LyC radiation in Haro 11:  the candidate presented in
this work via IPM, Knot A; the ULX and possible LLAGN, Knot B; or the
strong \lya\ source, Knot C. 

\acknowledgments
We thank Matt Brorby, Matthew Hayes, Genoveva Micheva, and G\"oran
\"Ostlin for useful discussions.  We are also grateful to the anonymous
referee for helpful comments.  This work was supported by NASA grant HST-GO-13702.

\end{document}